\documentclass[12pt]{article}
\usepackage{graphicx}
\usepackage{amssymb,amsmath,amsfonts,palatino,amsthm}
\usepackage{amssymb}
\usepackage{epstopdf}
\newcommand{\f}{\begin{equation}}
\newcommand{\ff}{\end{equation}}

\begin{document}

\title{Twistor relative locality }
\author{Lee Smolin\thanks{lsmolin@perimeterinstitute.ca} 
\\
\\
Perimeter Institute for Theoretical Physics,\\
31 Caroline Street North, Waterloo, Ontario N2J 2Y5, Canada}
\date{\today}
\maketitle

\begin{abstract}

We present a version of relative locality based on the geometry of twistor space.  This can also be thought of as a new kind of
deformation of twistor theory based on the construction of a bundle of twistor spaces over momentum space. Locality in space-time
is emergent and is deformed in a precise way when a connection on that bundle is non-flat.  This gives a precise and controlled
meaning to Penrose's hypothesis that quantum gravity effects will deform twistor space in such a way as to maintain causality
and relativistic invariance while weakening the notion that interactions take place at points in spacetime.  

\end{abstract}


\tableofcontents

\newpage

\section{Introduction}

Relative locality and twistor theory share the basic feature that space-time is derivative and is emergent from a more fundamental description of relativistic physics.  In twistor theory that more fundamental  geometry is twistor space, $\cal T$, whose points correspond, roughly, to  the trajectories of massless particles in  Minkowski space-time\cite{twistor}.  In relative locality\cite{rl,rl2} that more fundamental space is momentum space, $\cal P$.  

Interestingly enough, both share an aspiration, which is that the locality of interactions in space-time, being emergent, can be weakened by deforming the 
geometry of the space on which each theory is defined.  In each case this may be done without destroying the invariance of physics under the lorentz group.  
In relative locality this is done by  constructing a phase space which is a bundle of space times over momentum space, $\cal P$ and then deforming it by curving momentum space.    In twister theory this weakening of locality is done by deforming the complex geometry of twistor space.  

In each case these deformations of locality are tied to ideas about quantum gravity.  Penrose long ago hypothesized that when twistor theory was quantized casual relations would remain well defined but the coincidence or interaction of causal processes would no longer define points in space-time.  Relative locality, on the other hand, is defined as a phenomenological description of a new regime of a quantum theory of gravity in which $\hbar$ and $G$
are both considered negligible, while energies are probed comparable to their ratio, the Planck energy, $E_p = \sqrt{\hbar / G }$.

It is then natural to try to combine them.  Indeed twistor theory and relative locality have complementary weaknesses and strengths. Twistor theory emphasizes the causal relations amongst relativistic processes while relative locality emphasizes the primacy of energy and momentum and hence dynamics.  Each inverts the usual primacy of kinenamics  over dynamics.

Here we sketch one way to combine relative locality and twistor theory which was inspired by work on energetic causal sets\cite{first,second}.
These are causal sets\cite{cs,cs2} whose causal links carry energy and momentum which are conserved at events.  
There we defined what could be called a pre-geometry for relative locality in that space-time and the relative locality form of dynamics for
interacting relativistic particles emerges from a semiclassical approximation to the discrete processes.  In \cite{first} we found this
pre geometric theory could be elegantly expressed in terms of twistors.  The formulation that is discussed here follows naturally.

Relative locality has usually been described in terms of a phase space which is a cotangent bundle  constructed over momentum space.
Spacetime events emerge from interactions.  
Here we replace this with a bundle of twistor spaces over momentum space.  Events-when they emerge dynamically, 
are defined by the incidence  relation.  
A version of relative locality is then achieved by deforming a connection defined over this bundle.  The result is to give a precise
meaning to Penrose's idea that quantum gravity effects deform the notion of locality in space-time while preserving causality
and relativistic invariance.

The main result, a systematic weakening of locality, arises in a natural way.  Before the geometry of twistor space is deformed several massless particles interact at a point $z^{AA'}$ of Minkowski space-time which is the common solution to twistor incidence relations of the form,
\f
 \omega^D_I = z^{DA'}  \pi_{A'}^I
 \label{eq1}
\ff
where the components of the twistor representing the $I'th$ particle are $Z^\alpha_I = ( \omega^D_I, \pi_{A'}^I)$.  When 
twistor space is deformed these incidence relations are deformed to 
\f
 \omega^D_I = (z^{DA'} + \delta z^{DA'}_I )\pi_{A'}^I
 \label{eq2}
\ff
where the deviations from the original interaction event, $ \delta z^{DA'}_I $ are each different  functions of the connection of the bundle and the other momenta in the collision.  The $\delta z^{DA'}_I $ are proportional to $z^{AA'}$, realizing the idea that locality is relative\cite{rl,rl2}.   These deviations are proportional to the ratio of momenta to $E_p$  and thus represent the relative locality limit of a quantum theory of gravity, realized in twistor space.  

In the next section we study the undeformed theory, which gives a description of interacting particles in Minkowski space-time\footnote{The necessary background in twistor theory and relative locality can be gotten by reading the references \cite{twistor,rl,rl2}.}.  
The key step is the construction of an action for a system of interacting particles, expressed in twistor space.  In section 3
we deform the construction, leading to new versions of relative locality expressed by equations (\ref{eq1},\ref{eq2}).  Then in section 4 we derive the Poisson brackets from the action,
show that it leads to the standard twistor quantization, and propose that interactions be represented by constraints on the
space of twistor wavefunctions.

\section{The undeformed theory}

A point in twistor space, $\cal T$, is given by a two component spinor $\omega^A$ and a dual spinor $\pi_{A'}$.
\f
Z^\alpha = (\omega^A , \pi_{A'})
\ff
The $\pi_{A'}$ defines a point in momentum space $\cal P$ by the map ${\cal T} \rightarrow {\cal P}$
\f
p_{AA'}= \bar{\pi}_A \pi_{A'} 
\ff
These four-momenta are automatically null, so the constraint ${\cal H} $ is automatically implemented
\f
{\cal H}= \eta^{ab} p_a p_b =0
\ff

We consider a trajectory in twistor space
\f
Z^\alpha (s ) = (\omega^A  (s )  , \pi_{A'}  (s ) )
\ff
We will consider physical processes described as follows.  Free relativistic particles are described by paths in twistor space.  We will
see below that the trajectories of massless particles correspond to point in the projective twistor space $\cal PT$.
Interactions,  take place at events, where several paths exchange energy and momentum.  These events define points in space-time.  
At these events
four momenta are conserved.  This is at first enforced by a constraint at each event
\f
{\cal K}_{AA'} = \sum_I p_{AA'}^I
\label{K}
\ff
where the sum is over the particles that participate in each event.  Once we see how the action works we will be able to deform the
conservation law.

\subsection{The action}

Here is an action for a set of trajectories in twistor space, interacting at a set of events
\f
S= \sum_{paths} S^{path} + \sum_{events} S^{event} 
\ff
where the free action is
\f
S^{path} = \int_0^1 ds \left ( 
\imath \bar{\omega}^{A'} \dot{\pi}_{A'} - \imath  \omega^A \dot{\bar{\pi}}_A - \Omega ( \omega^A \bar{\pi}_A + \bar{\omega}^{A'} \pi_{A'}   -2S )
\right )
\label{Spath}
\ff
and the interaction action is
\f
S^{event}= z^{AA'} {\cal K}_{AA'} .
\ff
The integral in $S^{path}$ is between endpoints that define the beginning and end of a path.  Except for initial and final states the endpoints define
the $\pi_{A'}$ that participate in the interactions where the constraints (\ref{K}) are enforced.

The $z^{AA'}$ are lagrange multipliers, one for every event, that enforce the constraints (\ref{K}).  They live in a space dual to momentum space but to begin with
have no other role.

The $\Omega (s)$ are also Lagrange multipliers.  They enforce the constraints
\f
{\cal S} = \omega^A \bar{\pi}_A + \bar{\omega}^{A'} \pi_{A'}   -2S =0
\label{S}
\ff
that fix the helicity $S$ of the free particles.  This also defines the metric on twistor space
\f
Z^\alpha \bar{Z}^\beta g_{\alpha \beta } = \omega^A \bar{\pi}_A + \bar{\omega}^{A'} \pi_{A'} 
\ff

 Note that the term with time derivatives in the action (\ref{Spath}) differs from the form used in earlier works: $\int ds \bar{Z}_\alpha \dot{Z}^\alpha $, by an integration by parts.
 This is an indication that the interaction term and boundary term in the action break the conformal symmetry of $\cal T$ down to
 the lorentz group.

\subsection{Equations of motion}

Variations of $\omega^A (s) $ and $\bar{\omega}^{A'} (s)$ along the paths lead to the equations of motion
\f
\dot{\pi}_{A'} = -\imath \Omega \pi_{A'},\ \ \ \ \    \dot{\bar{\pi}}_{A} = \imath \Omega \bar{\pi}_{A}
\ff
which is solved by
\f
\pi_{A'} (s) = e^{-\imath  \Omega s } \pi_{A'}^0
\ff

The variations of $\pi_{A'} (s) $ and $\bar{\pi}_A (s)$ lead to two sets of equations.  An integration by parts yields an equation for each $s$ on each trajectory,
which gives the equations of motion for $\omega^A (s)$.
\f
\dot{\bar{\omega}}_{A'} =  \imath \Omega \bar{\omega}_{A'},\ \ \ \ \    \dot{\omega}_{A} = - \imath \Omega \omega_{A}
\ff
which is solved by
\f
\omega_{A} (s) = e^{-\imath  \Omega s } \omega_{A}^0
\ff
Hence 
\f
Z^\alpha (s) = e^{-\imath  \Omega s } Z^\alpha_0 
\label{solutions}
\ff 
so a path defines a circle of radius $2S$ in the twistor space (from eq. (\ref{S})) and hence a fixed point in {\it projective twistor space}.

There are also two equations from the variations of  $\pi_{A'} (s) $ and $\bar{\pi}_A (s)$ at each endpoint, getting contributions both
from the interaction action at each event and from the endpoints of the integrations by parts from the paths that meet
at that event.

These give the  incidence relations
\f
\omega^A = \imath z^{AA'} \pi_{A'},  
\label{teq}
\ff
\f
 \bar{\omega}^{A'} = -\imath z^{AA'} \bar{\pi}_{A}, 
 \label{teqbar}
 \ff
 
Initially we are in the twister space corresponding to the complexification of MInkowski space-time, so $\pi_{A'}$ and $\bar{\pi}_A$
are independent as are $\omega^A$ and $\bar{\omega}^{A'}$.  If we want to go to the real section we impose that the complex  conjugate
$\pi^*_{A} =\bar{\pi}_A$ and $\omega^{*A}  = \bar{\omega}^{A'}$.

On the real section (\ref{teq},\ref{teqbar}) together imply that $z^{AA'}$ are real
\f
{z}^{* A'A} = z^{AA'}
\label{reality}
\ff

But there are only real solutions to the incidence relations if $S=0$, so we only have real solutions for that case.

Note that we get one incidence relation for each path and each event at its endpoint.  Thus, suppose that
two paths given by twistors $Z^\alpha$ and $W^\alpha$ participate in an event, which means that their incidence relations
share a common solution $z^{AA'}$ to their incidence relations.  In twistor geometry this implies that
$Z^\alpha W_\alpha =0$.  

\section{Deforming twistor geometry}

In relative locality we study deformations of momentum space, $\cal P$.  Here we have to deform some aspect of
the geometry of twistor space.  What we will do is deform a connect that tells us how to parallel transport twistors
on momentum space.  

So long as we describe only massless particles there is no need to deform the metric of $\cal P$ or $\cal T$.

Here we do something 
else which is to construct a connection on a twistor bundle, $\cal B$, over momentum space, ${\cal P}$. That is to each $p_{AA'} \in {\cal P}$ we assign a space
of twistors, ${\cal T}_p = (\omega^A_p , \rho_{A' p})$.  Of course there is a special spinor at null momenta which is the 
$\rho_{A'}$ that yields $p_{AA'}$ through $p_{AA'} = \bar{\rho}_A \rho_{A'}$.  

Now given a momentum vector $p_{AA'}$ we define the left translation of a twistor at the origin of $\cal P$ to $p_{AA'}$ by
\f
L_p \pi_{A'}^0 = \pi_{A'}^p = \bar{U} [ p ]_{A'}^{B'} \pi_{B'}^0 ,\ \ \ \ \ \  L_p \omega_{A}^0 = \omega_{A}^p = U [ p ]_{A}^{B} \omega_B^0 
\ff

This defines a connection with group $SL(2,C) \times SL(2,C)$.

 Now given a null momentum $r_{AA'} = \bar{\rho}_A \rho_{A'}$ we can define
 \f
 p_{AA'} \oplus r_{AA' } = p_{AA'}  + U[ p ]_{AA'}^{BB'} r_{BB'} = p_{AA'}  + U[ p ]_{A}^{B}   \bar{U}[ p ]_{A'}^{B'}   \bar{\rho}_B \rho_{B'}.  
 \label{oplus}
 \ff
 Requiring that this take real momenta to real momenta implies that $\bar{U}$ is the complex conjugate of $U$, which restricts the
 connection to $SL(2,C)$.  
 
 Note that this is at first  an asymmetric relation because it is defined for any $p_{AA'}$ but only for null $r_{AA'}$.  On the other had the result
 $ p_{AA'} \oplus r_{AA' } $ is not generally null.  Hence there is a unique order for the triple and higher product
 \f
 p \oplus r \oplus q =  ( p \oplus r)  \oplus q
 \ff
so that we act always from the left.  Working this out we have
 \f
  ( p \oplus r)  \oplus q = (p + U [p] \cdot r ) \oplus q = p + U [p] \cdot r  +   U[p + U [p] \cdot r ] \cdot q 
 \ff
 If $p_{AA'}$ is small we can define connection coefficients and write
 \f
 U [ p ]_{A}^{B} = \delta_A^B + p_{CC'} \Gamma^{CC' B}_A  + \ldots, \ \ \ \   \bar{U} [ p ]_{A'}^{B'} = \delta_{A'}^{B'} + p_{CC'}\bar{\Gamma}^{CC' B'}_{A'} + \ldots . 
 \ff
  
  There is a natural choice for the connection $\Gamma^{CC' \  B}_A$ which is in terms of the relativistic Pauli matrices:
  \f
  \Gamma^{CC' \ B}_A = \frac{1}{M_p} [\sigma^{CC'}]^B_A
  \label{ansatz}
  \ff
 where $M_p$, the Planck mass sets the expected scale of deformations coming from quantum gravity.  These deform the dynamics without
 breaking the invariance of the theory under the lorentz group.  
 
Note that (\ref{oplus}) can be extended to an algebra over all the momentum space.  Even if $r_{AA'} \neq \bar{\rho}_{A} \rho_{A'}$ 
 we can define 
 \f
 p_{AA'} \oplus r_{AA' } =  p_{AA'}  + U[ p ]_{A}^{B}   \bar{U}[ p ]_{A'}^{B'}   r_{BB'}  
 \label{oplus2}
 \ff
 
 Thus,  a connection defined on a twistor bundle over $\cal P$ implies a deformation of the connection on $\cal P$ and hence a version of relative locality.

 One result is a deformation of the incidence relation which deforms locality.  As an example we may consider a trivalent vertex with
 \f
 {
\cal K}_{AA'} = ( p_{AA'} \oplus r_{AA'} ) \oplus q_{AA'}
 \ff
 where $p_{AA'},  r_{AA'}  $ and $ q_{AA'}$ are the four momenta of the three twistors representing three paths which interact at the event
 whose lagrange multiiplier is $z^{AA'}$.  These three twistors are represented by $(\omega^A, \pi_{A'}), (\lambda^A, \rho_{A'})$
 and $(\mu^A, \xi_{A'})$, respectively.

 We  find the  three  equation of motion for $\bar{\pi}_D ,   \bar{\rho}_D$ and $\bar{\xi}_D $.      These result in three deformed incidence relations,
 \begin{eqnarray}
 \omega^D &=& \imath z^{AA'} \frac{\partial {\cal K}_{AA'} }{\partial \bar{\pi}_D} 
 \\
 \lambda^D &=& \imath z^{AA'} \frac{\partial {\cal K}_{AA'} }{\partial \bar{\rho}_D} 
 \\
 \mu^D &=& \imath z^{AA'} \frac{\partial {\cal K}_{AA'} }{\partial \bar{\xi}_D} 
 \end{eqnarray}
 
 Writing out each to leading non-trivial order we find three equations that $z^{AA'}$ and the three twistors must simultaneously solve
  \begin{eqnarray}
 \omega^D 
&=& z^{DA'} \pi_{A'} 
\\
&+& z^{AA'} \pi_{C'} 
 \left (  \Gamma^{DC' B}_A (\bar{\rho}_B \rho_{A'} + \bar{\xi}_B \xi_{A'} )+ \bar{\Gamma}^{DC'B'}_{A'}( \bar{\rho}_A \rho_{B'}  + \bar{\xi}_A \xi_{B'}      )\right  )
 \nonumber
 \\  \nonumber
 \\
 \lambda^D &=& z^{DA'} \rho_{A'} 
\\
&+& z^{AA'}  
 \left (  \rho_{C'} ( \Gamma^{BC' B}_A  \bar{\xi}_B \xi_{A'} + \bar{\Gamma}^{BC'B'}_{A'}\bar{\xi}_A \xi_{B'}   )   
 + \bar{\pi}_C \pi_{C'} (\Gamma^{CC'D}_A \rho_{A'} + \bar{\Gamma}^{CC'B'}_{A'}   \rho_{B'}   \delta_D^A  )
 \right  )
 \nonumber
 \\  \nonumber
 \\
 \mu^D 
 &=& z^{DA'} \xi_{A'} 
\\
&+& z^{AA'}  
(    \bar{\pi}_C \pi_{C'}  +    \bar{\rho}_C \rho_{C'} )(\Gamma^{CC'D}_A \xi_{A'} + \bar{\Gamma}^{CC'B'}_{A'}   \xi_{B'}   \delta_D^A  )
 \nonumber
  \end{eqnarray}

These deformed incidence relations will have simultaneous solutions, but the three twistors involved will no longer be incident.  The
$z^{AA'}$ that solves these three equations will no longer be on any of the null lines defined by the three twistors.  
To see this, rewrite the preceding equations as
 \begin{eqnarray}
 \omega^D &=& (z^{DA'} + \delta z^{DA'}_1 )\pi_{A'} 
 \\
 \lambda^D &=&  (z^{DA'} + \delta z^{DA'}_2 )\ \rho_{A'} 
 \\
 \mu^D 
 &=&  (z^{DA'} + \delta z^{DA'}_3 )\ \xi_{A'} 
  \end{eqnarray}

where 
\begin{eqnarray}
\delta z_1^{DA'} &=&  z^{AE'} \left (  \Gamma^{DA' B}_A (\bar{\rho}_B \rho_{E'} + \bar{\xi}_B \xi_{E'} )+ \bar{\Gamma}^{DA'B'}_{E'}( \bar{\rho}_A \rho_{B'}  + \bar{\xi}_A \xi_{B'}      )\right  )
\\
\delta z_2^{DA'} &=& z^{AE'}  
 \left (  ( \Gamma^{BA' B}_A  \bar{\xi}_B \xi_{E'} + \bar{\Gamma}^{BC'B'}_{E'}\bar{\xi}_A \xi_{B'}   )   
 + \bar{\pi}_C \pi_{C'} (\Gamma^{CA'D}_A \rho_{E'} + \bar{\Gamma}^{CA'B'}_{E'}   \rho_{B'}   \delta_D^A  )
 \right  )
\\
\delta z_3^{DA'} &=& z^{AE'}  
(    \bar{\pi}_C \pi_{C'}  +    \bar{\rho}_C \rho_{C'} )(\Gamma^{CC'D}_A \delta^{A'}_{E'} + \bar{\Gamma}^{CC'A'}_{E'}  \delta_D^A  )
\end{eqnarray}

We see that $z^{AA'}$ is displaced from the three null lines, each by a different factor proportional to a product of $z^{AA'}$ and  $\Gamma^{CC'D}_A $, hence this is the typical distance between the
three null lines.  This gives meaning to the original idea of Penrose that twistor geometry-when deformed by quantum gravity effects, maintains
causal relations and relativistic invariance, while weakening the  conception that events take place at points of space-time.  
The fact that earh $\delta z^{AA'}$ is proportional to $z^{AA'}$ realizes the idea of relative locality that these deviations from
locality area always proportional to the displacement of the event from the observer.  Invariant quantitates are associated with closed loops of processes, while attribution of non-locality to specific events is only possible relative to each local observer.  
 
 \section{Twistor quantization}
 
 \subsection{Canonical theory}
 
 From the action (\ref{Spath}) we find the Poisson brackets
 \f
 \{ \pi_{A'} , \bar{\omega}^{B'} \} = -\imath \delta_{A'}^{B'} , \ \ \ \ \   \{ \bar{\pi}_{A} , {\omega}^{B} \} =  \imath \delta_{A}^{B} 
 \ff
 which is
 \f
  \{  Z^\alpha , \bar{Z}_\beta \} =  -  \imath \delta_{\alpha }^{\beta} ,  
 \ff
 making twistor space into the phase space of the theory.  Processes involving $N$ free or interacting particles have a phase space
 \f
 \Gamma = {\cal T}^N
 \ff

 The free action is totally constrained, indeed the constraint, $\cal S$ defined by (\ref{S}), which is unique, and hence first class,  generates gauge transformations,
 \f
 \delta_\Omega Z^\alpha = \Omega \{ Z^\alpha , {\cal S} \} = -\imath \Omega Z^\alpha 
 \label{deltaZ}
 \ff
Hence, we see that the solutions (\ref{solutions}) follow orbits of the gauge transformations, which can be interpreted for $\Omega$ real, as reparametrizations of the paths, reflecting the reparametrization invariance  of the free action (\ref{Spath}).

However the action (\ref{Spath}) is complex so 
(\ref{deltaZ}) can be taken as generating gauge transformations for $\Omega$ complex.  Hence the reduction from full twistor space ($C^4$) to projective
twistor space, ($CP^3$) is understood in this context as the reduction from the  kinematical to the physical phase space under a gauge invariance of
complex reparametrizations.

The extension to complex reparametrizations suggests an elegant extension to a string theory, which is
under development.

\subsection{Canonical quantization}
 
 We quantize in the representation where states are functionals of $Z^\alpha$.  The operators are
 \f
 \hat{Z}^\alpha \Psi ( Z )=  {Z}^\alpha \Psi ( Z ),  \ \ \ \  \hat{\bar{Z}}_\beta  \Psi ( Z )=  -\hbar \frac{\partial}{\partial Z^\beta } \Psi ( Z )
 \ff
 These realize the commutation relations
 \f
  [   \hat{Z}^\alpha , \hat{\bar{Z}}_\beta ] = \hbar  \delta_{\alpha }^{\beta} 
 \ff
 We impose the quantum constraint $\hat{\cal S}$ in the way natural for first class constraints:
 \f
0=  \hat{\cal S} \Psi (Z) = \left ( Z^\alpha  \frac{\partial}{\partial Z^\alpha }+ \frac{2S }{\hbar}  \right ) \Psi (Z)
 \ff
 which is the usual condition that twistor wave functions are homogeneous functions of fixed degree on $\cal T$.
 Note that the helicity $S$ could arise or be shifted by different operator orderings.
 
 Now consider a process involving $N$ free particles interacting at P events.  The quantum states are
 homogeneous functions of $N$ copies of $\cal T$,  $\Psi (Z_1 , \ldots, Z_N )$ 
 which are subject to an additional constraint for each event of the form
 \f
 \hat{\cal P}^I_{AA'} \Psi (Z_I ) =0
 \ff
 
 \subsection{Path integral quantization}
 
 Alternatively we can construct amplitudes for scattering processes using a path integral.  We consider amplitudes for
 scattering a fixed number of incoming particles, with fixed momenta and helicity to a similarly specified fixed set of
 outgoing particles.  This is the sum over amplitudes for all processes that can intermediate between the incoming and outgoing states.
 Each process is described by a diagram $\cal P$ where free particles interact at a set of events.  
 
 The amplitude for each diagram my be given by a path integral
 \f
 {\cal A}({\cal P}) = \int \prod_{particles} \prod_s d\omega^A (s) d \pi_{A'} (s) \Omega (s) \prod_{events} dz^{AA'} \delta (g.f.) \Delta_{FP}
 e^{\imath S}
 \ff
 where $g.f.$ refers to a gauge fixing of the reperametrization invariance (\ref{deltaZ}) and $\Delta_{FP}$ is the resulting
 Faddev-Poppov determinant.
 
 \subsection{Emergence of gravity}
 
 A feature of relative locality is that it is meant to describe a regime of quantum gravity phenomena in which $\hbar$ and $G$ have been taken 
 to zero together, while keeping their ratio $M_p^2 = \frac{\hbar}{G}$ fixed. In our formalism above $M_p$ comes in as the scale of
 the deformation, for example as in (\ref{ansatz})
 
  It follows from this that we cannot turn on $\hbar$ without also turning on $G$ as well,  because it will be defined from
 \f
 G=\frac{\hbar}{M_p^2}
 \ff
 
 If we use the quantum theory just defined to compute scattering processes there will be terms proportional to $\frac{\hbar}{M_p^2}$ and these
 will have gravitational strength.  The question is, will they describe gravity?

 \section*{Acknowledgements}

I am grateful to Andrzej Banburski, Linqing Chen, Marina Cortes and Laurent Freidel for conversations and feedback and to Lionel Mason for correspondence.
This research was supported in part by Perimeter Institute for Theoretical Physics. Research at Perimeter Institute is supported by the Government of Canada through Industry Canada and by the Province of Ontario through the Ministry of Research and Innovation. This research was also partly supported by grants from NSERC, FQXi and the John Templeton Foundation.


\end{document}